\documentclass{PoS}

\usepackage{graphicx}

\newcommand{\be}{\begin{equation}}
\newcommand{\ee}{\end{equation}\noindent}
\newcommand{\bea}{\begin{eqnarray}}
\newcommand{\eea}{\end{eqnarray}}
\newcommand{\nn}{\nonumber}

\newcommand{\maprightb}[1]{\smash{\mathop{
\hbox to 1cm{\rightarrowfill}}\limits_{#1}}}

\newcommand{\bc}{\begin{center}}
\newcommand{\ec}{\end{center}}

\newcommand{\matTwo}{\left(\begin{array}{rr}}
\newcommand{\matThree}{\left(\begin{array}{rrr}}
\newcommand{\emat}{\end{array}\right )}
\newcommand{\detTwo}{\left|\begin{array}{rr}}
\newcommand{\detThree}{\left|\begin{array}{rrr}}
\newcommand{\edet}{\end{array}\right |}

\title{Wilson fermions with imaginary chemical potential}

\ShortTitle{Wilson fermions with imaginary chemical potential}

\author{\speaker{Keitaro Nagata}\thanks{A footnote may follow.} \\
        Research Institute for Information Science and Education, 
    Hiroshima University,\\
  Higashi-Hiroshima 739-8527 JAPAN\\
        E-mail: \email{nagata@rcnp.osaka-u.ac.jp}}

\author{Atsushi Nakamura, Yoshiyuki Nakagawa, Shinji Motoki\\
Research Institute for Information Science and Education, 
Hiroshima University,\\
Higashi-Hiroshima 739-8527 JAPAN}
\author{Takuya Saito\\
Faculty of Science / Integrated Information Center, Kochi University, \\
Akebono-cho, Kochi 780-8520, JAPAN}
\author{Masatoshi Hamada \\
 Department of Physics, Kyushu University, \\
 Fukuoka 812-8581, JAPAN}

\abstract{We study the phase structure of the $N_f=2$ flavors QCD with 
imaginary chemical potential with the use of
the clover-improved Wilson quark action and renormalization-group improved 
gauge action. We calculate the Polyakov loop on $8^3\times 4$ lattice for 
$(\beta, \kappa) = (1.8, 0.1411), (1.9,0.1388), (1.95, 0.1377)$ 
with $\mu_I = 0.2618$. 
We find that the phase of the Polyakov loop shows a two-state signal 
indicating the first order phase transition. This transition occurs in the 
vicinity of  $\beta = 1.9$, which corresponds to the temperature 
$T/T_{pc}=  1.08$. 

We also present a reduction formula for the quark determinant of the Wilson fermion.
We discuss the feature of the matrix reduction formula.

}
\FullConference{The XXVII International Symposium on Lattice Field Theory - LAT2009\\
		 July 26-31 2009\\
		 Peking University, Beijing, China}

\begin{document}

\section{Introduction}
It is of prime interest to understand QCD at finite temperature 
and density. The lattice QCD with a non-zero baryon density
suffers from  a sign problem, where the determinant of the quark 
matrix becomes complex and Monte Carlo simulations become unavailable. 
Because of this, it is still challenge to understand the QCD at finite 
density. Several approaches have been investigated to overcome or circumvent 
the sign problem: multi-parameter reweighting, canonical approaches, 
imaginary chemical potentials, and so on. 

If the chemical potential is pure imaginary $\mu = \mu_R+ i \mu_I  
(\mu_R = 0)$, the quark determinant becomes real due to the identity 
$\Delta(\mu_I)^\dagger = \gamma_5 \Delta(\mu_I) \gamma_5$. 
Thus,  there is no sign problem in this case, and  the standard Monte Carlo 
algorithm can be applied to. The imaginary chemical potential was investigated 
with the use of the staggered (KS) fermion by  
De Forcrand and Philipsen~\cite{de Forcrand:2002ci} 
and D'Elia and Lombardo \cite{D'Elia:2002gd,D'Elia:2009qz}. Wu, Luo and Chen
considered the imaginary chemical potential using the standard 
Wilson quark action~\cite{Wu:2006su}. 

\begin{figure}[tbh]
\centering{
\includegraphics[width=5cm]{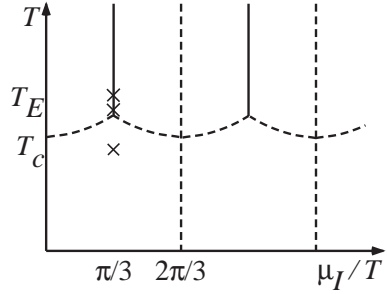}
}
\begin{minipage}{10cm}
\caption{A schematic figure of the phase structure in $(T, \mu_I/T)$ plane. 
Vertical solid lines show the critical lines of the Roberge-Weiss phase transition. 
Vertical dashed lines shows the Roberge-Weiss periodicity. 
The crossed symbols corresponds to points where the present simulations are 
performed.}\label{Nov0709fig1}
\end{minipage}
\end{figure}

The study of the phase structure in $(\beta,\mu_I)$-plane provides us with an 
understanding of the phase structure in $(\beta,\mu_R)$-plane through analytic 
continuation.
In the study of the imaginary chemical potential, it is important to consider 
the Roberge-Weiss periodicity~\cite{Roberge:1986mm}.  
In the presence of the quark, the $Z(3)$ symmetry is explicitly broken. 
However, if the chemical potential is pure imaginary, 
the $Z(3)$ symmetry is maintained via a translation of $\mu_I$ as 
\begin{eqnarray}
Z\left(T, \frac{\mu_I}{T}\right) = Z\left(T, \frac{\mu_I}{T} + \frac{2\pi}{N_c} k \right),
\end{eqnarray}
where  $k$ is an integer. Roberge and Weiss showed the existence of a phase 
transition at high temperature, where the phase of the Polyakov loop is an 
order parameter, while this phase transition does not occur at low 
temperature. A critical line corresponding to this phase transition is 
$\frac{\mu_I}{T} = \frac{\pi}{N_c}$ with a critical endpoint $T_E$, 
see Fig~\ref{Nov0709fig1}. 

In this study, we present two results.
First, we report a preliminary results on the study of the phase 
structure of the $N_f=2$ flavors QCD with an imaginary chemical potential by using 
the clover-improved Wilson quark action and renormalization-group improved 
gauge action on $8^3\times 4$ lattice for 
$(\beta, \kappa) = (1.8, 0.1411)$,  $(1.9,0.1388)$, and $ (1.95, 0.1377)$ 
with the imaginary chemical potential $\mu_I = 0.2618$. 
Second, we present a reduction formula for the quark determinant of the Wilson action, which 
is an extension of the reduction formula given in Ref~\cite{Gibbs:1986hi,Hasenfratz:1991ax,Fodor:2001pe}.

\section{Phase Structure with Imaginary Chemical Potential}

\subsection{Formulation}
We employ the renormalization group improved gauge action~\cite{Iwasaki:1985we}
\begin{eqnarray}
S_g = \frac{\beta}{6} \left[ c_0 \sum (1 \times 1 \; \mbox{loop}) 
+ c_1 \sum ( 1\times 2 \; \mbox{loop})\right], 
\end{eqnarray}
with $c_1 = -0.331$ and $c_0 = 1-8 c_1$ and clover-improved Wilson action, where
the quark matrix is written as 
\begin{eqnarray}
\Delta(x,y) =  \delta_{x, x^\prime} 
&-&\kappa \sum_{i=1}^{3} \left[
(1-\gamma_i) U_i(x) \delta_{x^\prime, x+\hat{i}} 
+ (1+\gamma_i) U_i^\dagger(x^\prime) \delta_{x^\prime, x-\hat{i}}\right] \nonumber \\
 &-&\kappa \left[ e^{+\mu} (1-\gamma_4) U_4(x) \delta_{x^\prime, x+\hat{4}}
+e^{-\mu} (1+\gamma_4) U^\dagger_4(x^\prime) \delta_{x^\prime, x-\hat{4}}\right] \nonumber \\
&-& \delta_{x, x^\prime} C_{SW} \kappa \sum_{\mu \le \nu} \sigma_{\mu\nu} 
F_{\mu\nu}.
\end{eqnarray}
For the coefficient of the clover term $C_{SW}$, 
we use a result obtained in the one-loop perturbation theory~\cite{Sheikholeslami:1985ij},  $C_{SW} = ( 1- 0.8412 \beta^{-1})^{-3/4}$. 
We calculate the Polyakov loop $\langle L\rangle = | L | \exp( i \phi)$, where 
the phase $\phi$ is an order parameter of the Roberge-Weiss phase transition~\cite{Roberge:1986mm}.

\subsection{Results}

We perform simulations on $8^3\times 4$ lattice by using the standard hybrid Monte Carlo algorithm. 
In the present study,  we carry out the simulations for three set of 
parameters $(\beta, \kappa) = (1.8, 0.1411)$, $(1.9,0.1388)$, and  $(1.95, 0.1377)$.  
All for these three cases, we use a fixed value of the imaginary chemical potential 
$\mu_I = 0.2618$, which is on the line of the Roberge-Weiss phase transition $\mu_I / T= \pi / N_c $. 
These parameters correspond to lines of the constant physics with $m_{PS}/m_V=0.8$ in 
the absence of the chemical potential~\cite{Maezawa:2007fc}. 
We set a step size of the molecular dynamics to be $\delta \tau = 0.02$ and 
the number of the molecular dynamics to be 50, which gives 
the length of a molecular dynamics trajectory to be one.  We generate 10, 000 trajectories and 
measure the Polyakov loop for each trajectory.
\begin{figure}[tbh]
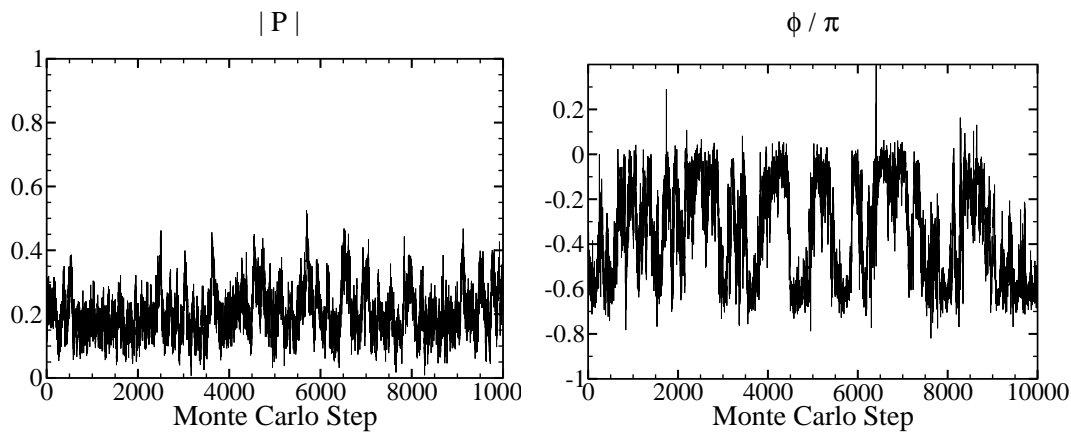

\centering{
\includegraphics[width=7cm]{outb1.9m2617hots100to309.pol.abs.eps}
\includegraphics[width=7cm]{outb1.9m2617hots100to309.pol.arg.eps}}
\begin{minipage}{11cm}
\caption{Monte Carlo History of the Polyakov loop for $(\beta, \kappa, \mu_I) 
= ( 1.9, 0.1388, 2.618)$. The left panel shows the absolute value $|P|$ and 
the right panel shows the phase $\phi$.}
\label{Oct1509fig1}
\end{minipage}
\end{figure}
\begin{figure}[tbh]
\centering{
\includegraphics[width=7cm]{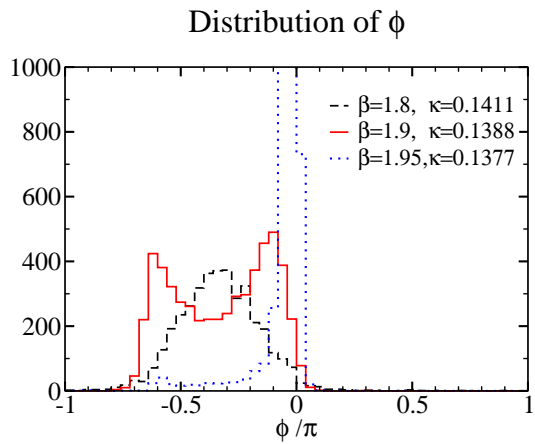}}
\begin{minipage}{11cm}
\caption{A histogram of the distribution of the phase of the Polyakov loop $\phi$.
The dashed (black), solid (red) and dotted (blue) lines correspond to 
$(\beta, \kappa) = (1.9, 0.1411), (1.9, 0.1388)$ and $(1.95, 0.1377)$, respectively. 
We use the same value of the imaginary chemical potential $\mu_I = 0.2618$ for 
these three sets of the parameters. }
\label{Oct1509fig2}
\end{minipage}
\end{figure}

Figures~\ref{Oct1509fig1} show the history of the Monte Carlo simulations 
for the absolute value and the phase of the Polyakov loop for $\beta=1.9$. 
The phase of the Polyakov loop $\phi$ shows the two-state structure with $\phi = 0$ and $-\pi /3$. 
This indicates the first order Roberge-Weiss phase transition. 

Figure~\ref{Oct1509fig2} shows the distribution of the phase of the 
Polyakov loop $\phi$, where the first 5,000 steps are removed as thermalization. 
Following  Ref.~\cite{Maezawa:2007fc}, three sets of the parameters correspond to the 
temperature $T/ T_{pc} = 0.93$ for $(\beta, \kappa) = (1.8, 0.1411)$, 
$T/T_{pc} = 1.08$ for $(1.9,0.1388)$ and $T/T_{pc} = 1.20$ for $ (1.95, 0.1377)$, 
where $T_{pc}$ is a pseudocritical temperature.
It follows from the figures that at low temperature $(\beta = 1.8)$, 
the system has one vacuum with the value of the order parameter $\phi \sim 0.3$.
As temperature becomes higher ($\beta=1.9$), two-state behavior appears with 
the two vacua $\phi =0 $ and $\phi = - \pi /3 $, although the peak structure is not sharp
due to the small lattice size. As temperature becomes much higher $(\beta=1.95)$, 
the vacuum shift to $\phi = 0$. 
We find the onset of the Roberge-Weiss phase transition is  
in the vicinity of $\beta = 1.9$ for which $T/T_{pc} = 1.08$. This is almost 
consistent with the previous result obtained by the standard Wilson quark 
action~\cite{Wu:2006su}. On the other hand, the phenomenological 
approaches~\cite{Sakai:2008um, Sakai:2008py, Kouno:2009bm, Braun:2009gm}
reported larger values $T_E = 200 \sim 210$ MeV.  
This dicrepancy is probably caused by the fact that we employed large quark mass 
$m_{ps}/m_V \sim 0.8$. 

\section{Gibbs formula for Wilson fermions}

In this section, we derive a formula which is indispensable if we use
Wilson fermions for 
the multiparameter reweighting by Fodor and Katz (i.e.,
using no Taylor expansion)~\cite{Fodor:2001pe}, and also for the canonical
expression. Danzer, Gattringer and Liptak studied a decomposition of the 
quark determinant for Wilson action\cite{Danzer:2008xs,Danzer:2009sr}. 
Here we derive an alternative expression.

We can write Wilson fermion matrix as
\be
\Delta = B -  z^{-1} \kappa (r-\gamma_4) V - z \kappa (r+\gamma_4) V^\dagger , 
\ee
where
\bea
B(x,x') &\equiv&  \delta_{x,x'}
 - \kappa \sum_{i=1}^{3} \left\{
        (r-\gamma_i) U_i(x) \delta_{x',x+\hat{i}}
      + (r+\gamma_i) U_i^{\dagger}(x') \delta_{x',x-\hat{i}} \right\}
\nn \\
&+& (Clover) , 
\eea
and
\be
V(x,x') \equiv 
 U_4(x) \delta_{x',x+\hat{4}}.
\ee
We introduce $z$ as
$
z \equiv e^{-\mu} . 
$
Now we rewrite $\Delta$:
\bea
\det \Delta &=& \det (B -  z^{-1} \kappa (r-\gamma_4) V - z \kappa (r+\gamma_4) V^\dagger), \nn \\
&=&
z^{-N} 
\left|
\begin{array}{cccc}
 & & &
 \\
-BV-z(-\kappa(r+\gamma_4)) & & I &
\\
 & & &
\\
  \kappa(r-\gamma_4)V^2       & & -z &
\\
 & & &
\end{array}
\right|
/\det V\; ,  \nn
\\
&=&
z^{-N} 
\left|
\left(
\begin{array}{cc}
-BV         & I
\\
\kappa(r-\gamma_4)V^2    & 0
\end{array}
\right)
- z
\left(
\begin{array}{cc}
-\kappa(r+\gamma_4)  &  0
\\
  0      &   I
\end{array}
\right)
\right|_.
\eea
Here $N$ is a rank of the block matrices, such as $B$ and $V$, 
$N\equiv N_c\times 4\times N_x N_y N_z N_t$.
By exchanging the columns and raws, this matrix now reads 
\be
\det\Delta = z^{-N} \det(T-zS).
\label{GenEigenValueProblem}
\ee
Here we describe the matrices $T$ and $S$ as block matrices in time-plane, 
\bea
T
=
\left(
\begin{array}{c|c|ccc|c}
   0 &  t_1 & 0 & \cdots& &  0 
\\ \hline
   0  & 0 & t_2  &\cdots  &  &  0 
\\ \hline
  0 & 0 & 0 & \cdots & & 
\\ 
 \cdots  &\cdots  &\cdots & \cdots& \cdots  &\cdots
\\ 
   &  & &\cdots    & t_{N_t-2} & 0 
\\ \hline
    0    & 0 & &  \cdots & 0 &  t_{N_t-1} 
\\ \hline
t_{N_t} & 0 &  &\cdots &0&  0 \\
\end{array}
\right) ,\; \; S = 
\left(
\begin{array}{c|c|ccc|c}
   s & 0  & 0 & \cdots& &  0 
\\ \hline
   0  & s & 0  &\cdots  &  &  0 
\\ \hline
  0 & 0 & 0 & \cdots & & 
\\ 
 \cdots  &\cdots  &\cdots & \cdots& \cdots  &\cdots
\\ \hline
    0    & 0 & &  \cdots & s &  0 
\\ \hline
   0 & 0 &  &\cdots &0&  s \\
\end{array}
\right), 
\nonumber
\eea
where 
\be
t_i = 
\matThree
-B_i V_{i,i+1}                        &  & 1
\\
\kappa (r-\gamma_4)V_{i-1,i}V_{i,i+1} &  & 0
\emat, \; \;
s = 
\left(
\begin{array}{cc}
-\kappa(r+\gamma_4)  &  0
\\
  0      &   I
\end{array}
\right) .
\ee
Each $t_i$ and $s$ is $(4N_cN_xN_yN_z)\times(4N_cN_xN_yN_z)$
matrix.

Even though the matrix $S$ does not have an inverse,
the formula (\ref{GenEigenValueProblem}) can be evaluated by the
generalized eigen value problem,
\be
T\vec{X} = z S \vec{X} .
\ee
Namely the generalized Schur decomposition\cite{Matrix} tells us that 
there exist unitary $Q$ and $Z$ such that $Q^\dagger S Z$ and $Q^\dagger T Z$
are upper triangular.  We write their diagonal elements, ${\alpha_i}$ and
$\beta_i$, respectively. Then
\be
\det (T-zS) = \det(Q Z^\dagger) \prod _i (\alpha_i - z \beta_i).
\ee
This is a Gibbs formula for Wilson fermions.  However, we evaluated ${\alpha_i}$
and $\beta_i$ using LAPACK routine, and there is an accuracy problem.
Then we go further to obtain a satisfactory formula.

Let us rewrite the determinant of $T-zS$ as
\bea
&&\det(T-zS)
=
\left|
\begin{array}{c|c|ccc|c}
t_{N_t} & 0 &  &\cdots & 0 &  -zs 
\\ \hline
   -zs &  t_1 & 0 & \cdots& &  0 
\\ \hline
   0  & -zs & t_2  &\cdots  &  &  0 
\\ \hline
  0 & 0 & -zs & \cdots & & 
\\ 
 \cdots  &\cdots  &\cdots & \cdots& \cdots  &\cdots
\\ 
   &  & &\cdots    & t_{N_t-2} & 0 
\\ \hline
    0    & 0 & &  \cdots & -zs &  t_{N_t-1} 
\end{array}
\right| .
\eea
We multiply  matrices $Q_1$ and $Q_2$ from the right
\be
Q_1 = \left(
\begin{array}{c|c|cc|c}
I & 0   &  &\cdots  & t_{N_t}^{-1}\, zs 
\\ \hline
0 & I   & 0 & \cdots&   0 
\\ \hline
0  & 0  & I  &\cdots  &    0 
\\ 
 \cdots  &\cdots  &\cdots & \cdots  &\cdots
\\ 
   &  & &\cdots     & 0 
\\ \hline
    0    & 0 &  \cdots &  & I 
\end{array}
\right), \; \; 
Q_2 = \left(
\begin{array}{c|c|cc|c}
I & 0   &  &\cdots  & t_1^{-1}\, zs\, t_{N_t}^{-1}\, zs 
\\ \hline
0 & I   & 0 & \cdots&   0 
\\ \hline
0  & 0  & I  &\cdots  &    0 
\\ 
 \cdots  &\cdots  &\cdots & \cdots  &\cdots
\\ 
   &  & &\cdots     & 0 
\\ \hline
    0    & 0 &  \cdots &  & I 
\end{array}
\right).
\ee
Here $\det Q_1 = \det Q_2 = 1$. Then, we obtain 
\bea
\det(T-zS) &=& \det (T-zS) Q_1 Q_2 , 
\nn \\
&=& |t_{N_t}|
\nn \\
\times 
&& \left|
\begin{array}{c|c|cc|c}
    t_1 & 0 & \cdots& &  0 
\\ \hline
    -zs & t_2  &\cdots  &  & -zs\,t_1^{-1} zs\, t_{N_t}^{-1}\, zs  
\\ \hline
   0 & -zs & \cdots & & 
\\ 
 \cdots  &\cdots & \cdots& \cdots  &\cdots 
\\ 
   & &\cdots    & t_{N_t-2} & 0 
\\ \hline
  0 & &  \cdots & -zs &  t_{N_t-1} 
\end{array}
\right| , 
\nn \\
&=& |t_{N_t}|\times |t_1|\times
\nn \\
&& \left|
\begin{array}{c|c c|c}
     t_2  &\cdots  &  & -zs\,t_1^{-1} zs\, t_{N_t}^{-1}\, zs  
\\ \hline
    -zs & \cdots & & 
\\ 
 &\cdots & \cdots &\cdots 
\\ 
  &\cdots    & t_{N_t-2} & 0 
\\ \hline
 0  &  \cdots & -zs &  t_{N_t-1} 
\end{array}
\right|.
\eea
We go further recursively.
\bea
\det(T-zS) &=& \det (T-zS) Q_1 Q_2 \cdots Q_{N_t-2}, 
\nn \\
\nn \\
&=& |t_{N_t}|\times |t_1|\times
\cdots |t_{N_t-3}|
\nn \\
\times
&&
\begin{array}{|cccc|}
t_{N_t-2}  &&&  -zs\,t_{N_t-3}^{-1}\,zs \cdots t_1^{-1}\,zs\,t_{N_t}^{-1}\,zs
\\
-zs        &&& t_{N_t-1} 
\end{array} \; .
\eea

Since
\bea
&&
\left(
\begin{array}{cccc}
t_{N_t-2}  &&&  -zs\,t_{N_t-3}^{-1}\,zs \cdots t_1^{-1}\,zs\,t_{N_t}^{-1}\,zs
\\
-zs        &&& t_{N_t-1} 
\end{array}
\right)
=
\nn \\
&&
\matTwo
t_{N_t-2} & 0
\\
0         & I
\emat
\times 
\left(
\begin{array}{cccc}
I  &&&  -t_{N_t-2}^{-1}\,zs\,t_{N_t-3}^{-1}\,zs 
\cdots t_1^{-1}\,zs\,t_{N_t}^{-1}\,zs
\\
-zs        &&& t_{N_t-1} 
\end{array}
\right) , 
\eea
we obtain
\bea
\det(T-zS) &=& 
|t_{N_t}|\times |t_1|\times
\cdots |t_{N_t-2}|
\nn \\
\times &&
|t_{N_t-1} 
- zs\,t_{N_t-2}^{-1}\,zs\,t_{N_t-3}^{-1}\,zs 
\cdots t_1 \,zs\,t_{N_t}^{-1}\,zs
| , 
\nn \\
&=&
|t_{N_t}|\times |t_1|\times \cdots |t_{N_t-2}|\times |t_{N_t-1}|
\nn \\
\times 
&&
|I 
- t_{N_t-1}^{-1}\,zs\,t_{N_t-2}^{-1}\,zs\,t_{N_t-3}^{-1}\,zs 
\cdots t_1 \,zs\,t_{N_t}^{-1}\,zs
|, 
\nn \\
&=&
|P| \times
| I - z^{N_t}
 t_{N_t-1}^{-1}\,s\,t_{N_t-2}^{-1}\,s\,t_{N_t-3}^{-1}\,s 
\cdots t_1 \,s\,t_{N_t}^{-1}\,s
| , 
\eea
where $P = t_1 t_2 \cdots t_{N_t}$. 

Therefore, the ratio of the fermion determinant with $z=e^{-\mu}$ to that with
$z=1$, i.e., $\mu=0$ is
\be
\frac{\det\Delta(\mu)}{\det\Delta(\mu=0)}
= z^{-N} 
\frac{
\det\left( I - z^{N_t} Q \right) 
}{ \det\left( I - Q \right) 
} . 
\ee
Here $Q$ is a matrix of $L\times L$ with $L\equiv 2 \times 4\times N_c N_xN_yN_z$ 
and is given as
\be
Q \equiv t_{N_t-1}^{-1}\,s\,t_{N_t-2}^{-1}\,s\,t_{N_t-3}^{-1}\,s
\cdots t_1 \,s\,t_{N_t}^{-1}\,s . 
\ee
If we can diagonalize $Q \to \mbox{diag}\{q_1, q_2, \cdots , q_L\}$,
then
\be
\frac{\det\Delta(\mu)}{\det\Delta(\mu=0)}
= z^{-N} 
\frac{
\prod_{l=1}^{L} (1-z^{N_t} q_l) 
}{
\prod_{l=1}^{L} (1-q_l) 
} .
\ee
Although it contains inverse matrix calculations, $t^{-1}$, the matrix $Q$ 
does not depend on $N_t$.

\section{Summary}

We have investigated the phase structure of the QCD with the imaginary 
chemical potentials. 
We employed $N_f=2$ flavors QCD with the renormalization-group improved gauge 
action and the clover-improved Wilson quark action. 
We performed simulations on $8^3\times 4$ lattice for $(\beta, \kappa) 
= (1.8, 0.1411), (1.9,0.1388), (1.95, 0.1377)$ with $\mu_I = 0.2618$. 
We found that the phase of the Polyakov loop showed the two-state signals 
at $(\beta, \kappa)=(1.9,0.1388)$, which indicates the first order phase 
transition. This phase transition occurs in the vicinity of $\beta = 1.9$, 
which corresponds to $T/T_{pc}= 1.08$.
In order to understand the phase structure of the QCD, such as the pseudocritical 
line $\beta_c(\mu)$ and the endpoint of the Roberge-Weiss phase transition,  
the simulation for other values of $(\beta, \kappa, \mu_I)$ is in progress. 

AN is supported by the Grant for Scientific Research [(C) No.20340055]
from the Ministry of Education, Culture, Science and Technology, Japan.

\end{document}